          [1999/12/02 1.01 (PWD)]
\documentclass[a4paper,twocolumn]{esapub} 
\usepackage{natbib,graphicx}

\title{Testing the origin of the gamma-ray background}
\author{Andrzej M. So\l tan}
\affil{Nicolaus Copernicus Astronomical Center, Bartycka 18, 00-716 Warsaw, Poland}
\author{Rados\l aw Marcinkowski}
\affil{Physics Department, Warsaw University, Ho\.za 69, 00-681 Warsaw, Poland}

\begin{document}


\maketitle

\begin{abstract}
Fluctuations of the gamma-ray background (GRB) are investigated.  It is
assumed that the GRB is produced by discrete sources. Two basic classes
of objects as dominant contributors to the GRB are considered: a) blazars
and b) starburst galaxies. Predicted $\log N - \log S$ counts are calculated
for these two models. In both cases strong cosmic evolution is needed
to produce a substantial fraction of the observed GRB. Since on the
average blazars
are several orders of magnitude more luminous than starburst galaxies,
their number required to saturate the GRB is much smaller than the number
of starburst galaxies. In effect, fluctuations of the GRB in the blazar
model are relatively high. It is shown that moderate improvement in the
measurement accuracy of the GRB flux will allow us to discriminate between
these models.
\end{abstract}

\section{Introduction}

Using the present detection techniques we are unable to define a true nature
of the diffuse extragalactic gamma-ray background. The status of the GRB
is now analogous to that of the X-ray backgrund in 1960s and 70s. Although
some classes of objects are known $\gamma$-ray emitters, it is by no means
certain what fraction of the GRB is produced by any particular classes of
objects and even by the discrete sources in general.
Furthermore, it seems likely that some fraction of the diffuse emission
considered previously to be of the extragalactic origin is produced in
the halo of our Galaxy \citep{strong99}. \cite{dar00} postulate that most
if not all of the GRB is produced in the Galactic halo.

One should expect that structural features of the GRB are related to its origin.
It is likely that trully diffuse processes produce the GRB of smooth distribution
without strong well defined fluctuations. On the contrary, small scale
variations would indicate the discrete nature of the background.
In the present paper we concentrate on a class of models which assume that
the whole GRB is produced by a well defined population of discrete sources.
We discuss two cases in detail.
The source counts are estimated and prospects for the detection
of individual sources by the present-day and the near future instruments are
assessed. Next we estimate amplitude of the GRB fluctuations in various
angular scales characteristic to both these models.

\section{Discrete source model}

It is assumed that the extragalactic $\gamma$-ray background is produced
by a population of unresolved discrete sources. We assume further that these
sources are distributed randomly, i.e. effects of source clustering on the
GRB fluctuations are neglected (it is straightforward to show that for
any realistic model of the source clustering the latter assumption
is reasonable). Since it is not clear which class of sources
contributes most to the GRB, we consider two types of objects which potentially
could generate the GRB: {\it a)}  blazars and {\it b)} starburst galaxies.
Blazars are known
to be $\gamma$-ray emitters (e.g. \citealt{mukherjee97}), while starburst
have been considered as potential sources (e.g. \citealt{soltan99}).
The low sensitivity and poor angular resolution of the present-day
$\gamma$-ray telescopes allows only for the detection of the most
luminous objects, mostly blazars. However, the contribution of blazars to
the GRB remains highly uncertain and depends strongly on the cosmic evolution.
The question whether the GRB could be produced by the starburst galaxies
also cannot be answered directly. Locally these objects are much weaker
than blazars, but their space density is higher and there are indications
that these objects also are subject to the strong cosmic evolution
(\citealt{saunders90}, \citealt{pearson96}).

\section{Source counts}

The source counts are related to the luminosity function (LF) by the formula:
\begin{equation}
N(>\!S) = \int_{z_{min}}^{z_{max}} dz\,{dV\over dz}\,\int_{L_{min(z,S)}}^{L_{max}}
            \phi(L,z)\,dL,
\end{equation}
where $S$ and $L$ denote the $\gamma$-ray flux and luminosity, respectively,
$\phi(L,z)$ is the evolving with redshift luminosity function and all the
symbols have their ususal meaning. Low luminosty limit in the second intergral
is defined as $L_{min(z,S)} = 4\pi S D_L^2 (1+z)^{-(1+\alpha)}$,
where $D_L$ is the luminosity distance and $\alpha$ is the average
energy ($E$) spectral index of the $\gamma$-ray sources: $L(E)\sim E^{\alpha}$.
In the subsequent calculations data taken from the literature
have been scaled to the Hubble constant
$H_{\circ} = 100$\,km\,s$^{-1}$Mpc$^{-1}$ and $q_{\circ}= 0$.

\subsection{AGN model}

Predicted $\log N - \log S$ counts of the $\gamma$-ray blazars have been
calculated using data on the local luminosity function and the
luminosity evolution rate of these objects from \cite{chiang98}.
AGNs loud in the $\gamma$-rays detected by EGRET have been used.
The local LF shows pronounced flattening at luminosities below
$\sim 10^{46}$\,erg\,s$^{-1}$. Estimates of the luminosity evolution
parametrized by a power-law $(1+z)^{\beta}$ give $\beta = 2.7$ at redshifts
smaller than 2.5. At higher redshifts the evolution slows down and in the
present calculations we assumed $\beta = 0$. Although both the LF and
the evolution are subject to large uncertainties, the overall shape
of the $\log N - \log S$ predicted by the model only weakly depends
on the particlular choice of parameters. The counts shown in Fig.~\ref{lognlogs}
are labelled 'AGN'. According to the present model counts flatten substantially
at fluxes of $\sim 10^{-8}\,{\rm ph/(s\,cm^2)}$. It implies that most of
the GRB is produced by relatively small number of bright sources.

\subsection{Starburst galaxy model}

Significantly different predictions concerning the source counts and the
structure of the GRB are obtained in the model of the $\gamma$-ray emission
by starburst galaxies proposed by \cite{soltan99}. In this case the
$\gamma$-rays are produced as a result of the inverse Compton scattering
of the far infrared (FIR) photons on the cosmic ray electrons. Both cosmic
microwave background photons and intrinsic galaxy radiation contribute
to the production of $\gamma$-rays. Since cosmic ray electrons manifest
their presence in the galaxy via synchrotron radiation, measurements of
radio luminosities of starburst galaxies provide data on the number
of these electrons. Consequently, the correlation between the FIR and radio
luminosities observed for the starburst galaxies (e.g. \citealt{chi90})
allows for relatively safe estimates of the $\gamma$-ray luminosity
of each galaxy and the local volume emissivity of the whole starburst
galaxy population (see \cite{soltan99} for details). In fact, the obtained
figures are just lower
limits for the high energy emission of these objects. This is because
apart from the inverse Compton mechanism, one should also take into
consideration Bremsstrahlung emitted by cosmic ray electrons in the
interaction with the galactic gas.

\begin{figure}
\centering
\vspace{-10mm}
\includegraphics[width=0.8\linewidth]{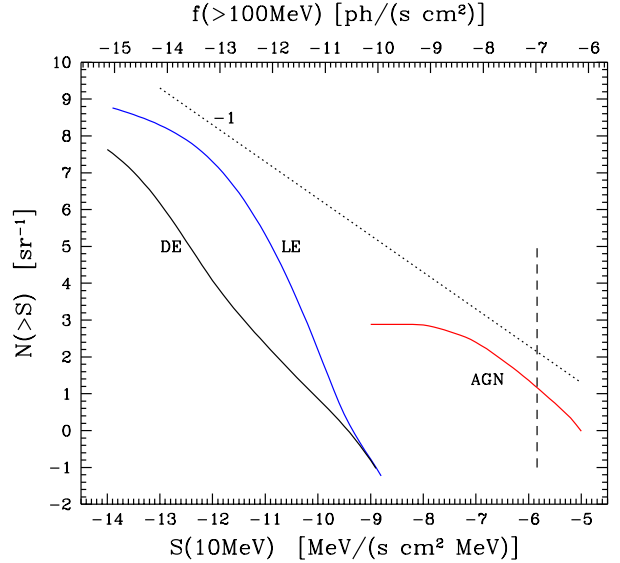}
\caption{Counts of the $\gamma$-ray~sources producing
the GRB. Energy flux of the lower abscissa axis and the photon flux of the
upper axis are scaled assuming the average photon index of -2.1.
The curve labelled 'AGN' shows the $\log N - \log S$ counts
if the GRB is dominated by blazars; curves 'DE' and 'LE" represent counts
predicted in the starburst galaxies model with the {\it density evolution}
and {\it luminosity evolution}, respectively. Verical line marks typical
detection threshold of EGRET; source counts which have common envelope
line with a slope of $-1$\, produce similar fraction of the GRB.\label{lognlogs}}
\end{figure}

The FIR evolution rate of starburst galaxies is not well determined,
but some direct estimates at low redshifts  are available
(e.g. \citealt{saunders90}). Various observations provide data on
the star formation history covering redshifts up tu $\sim 4$ (e.g.
\citealt{moorwood96}). There is an agreement that star formation rate (SFR)
below $z = 1 - 1.5$ varies roughly as $(1+z)^{3.5}$, and for higher $z$
is either stable or smoothly decayes to low levels at $z > 4$
(e.g. \citealt{franceschini97}, \citealt{hopkins00}).
Approximate data on SFR do not constrain strongly our estimates of the
evolution of the $\gamma$-ray volume emissivity and source counts.
This is because the $\gamma$-ray source counts depend strongly on the
evolution type of the FIR properties of starburst galaxies. If these objects
are subject to the density evolution (DE), the normalization of the FIR LF
increases with redshift, but luminosities on
the average do not evolve. In the luminosity evolution (LE)
case, the space concentration of objects is constant while their FIR
luminosities ($L_{FIR}$) increase with redshift.

High energy luminosity of the individual galaxy due to the inverse Compton
effect is proportional to the total number of cosmic ray electrons and 
the energy density of FIR photons (e.g. \citealt{rybicki79}). 
Since production of the cosmic ray electrons is also related to the $L_{FIR}$
(indirectly via rate of supernova explosions), the $\gamma$-ray luminosity,
$L_{\gamma}$, should be proportional roughly to the square
of the $L_{FIR}$. A more scrupulous analysis of the relevant correlations
between radio luminosity, cosmic ray electron spectrum and strength
of the magnetic field, $B$, as well as radio and FIR luminosities,
and the correlation between the $B$ and $L_{FIR}$ (\citealt{chi90}) gives
the number of electrons proportional to $L_{FIR}^{0.95}$ (\citealt{soltan99}),
and finally $L_{\gamma} \sim L_{FIR}^{1.95}$.
Thus, in the case of the LE of FIR luminosity function, we expect
a much stronger growth of the $\gamma$-ray emissivity by the starburst
galaxies than in the DE case. Accordingly, the predicted counts for
the LE model are much steeper. In Fig.~\ref{lognlogs} curves labelled 'LE'
and 'DE' show the count for both models (in both cases evolution rate
of the FIR volume emissivity is the same).

\section{Fluctuations}

\cite{sreekumar98} using the EGRET data above 100\,MeV have
estimated the isotropy of the GRB by measuring the backgound intensity
over the sky (far from the galactic center and the galactic plane).
Analysis of 28 regions of approximately 0.26 sr each
gave the average flux of $1.36\times
10^{-5}$\,ph\,(s\,cm$^2$\,sr)$^{-1}$ with the rms scatter between
the regions of $0.23\times 10^{-5}$\,ph\,(s\,cm$^2$\,sr)$^{-1}$.
The average uncertainty of the measurement for each field was equal to
$0.21\times 10^{-5}$\,ph\,(s\,cm$^2$\,sr)$^{-1}$. This apparent concordance
implies that
the detected fluctuations are consistent with the noise of the
EGRET measurements. Below we estimate amplitude of the GRB fluctuations
generated in the present models and compare our predictions with the
parameters of the INTEGRAL instruments.

Although integrated fluxes of the GRB predicted by the 'AGN' and
'starburst LE' models are similar, the source
counts in these two cases are distinctly different. The vertical dashed line
shows the threshold sensitivity of EGRET ($\sim 10^{-7}\,{\rm ph/(s\,cm^2)}$
above 100 MeV). A number of sources above this threshold have
been in fact detcted by EGRET. 
However, these sources produce a small fraction of the total GRB.
In the 'starburst LE' model most of the GRB is produced by sources fainter
by several orders of magnitude, while in the 'AGN' model the substantial
fraction of the GRB is generated by sources with
$f > 10^{-8}\,{\rm ph/(s\,cm^2)}$, i.e. just an order of magnitude weaker
than the EGRET limit. Thus,
with the sensitivity improved by an order of magnitude below the EGRET
threshold, both models would be distinguished directly.
Since the predicted sensitivities of the main INTEGRAL instruments,
IBIS and SPI, are comparable to EGRET (assuming the average source spectral
slope of $-2.1$), it is unlikely that dicrete
source observations would allow an unambiguous differentaition of both
models. However, one can assess the source counts
below the threshold of $S(10\,{\rm MeV}) = 10^{-6}\,{\rm MeV/(s\,cm^2\,MeV)}$
(what roughly corresponds to  $f(>\!100\,{\rm MeV}) = 10^{-7}\,{\rm ph/(s\,cm^2)}$)
using the fluctuation analysis.
The number of sources per steradian producing most of the GRB in the AGN
model is relatively small. In this case fluctuation of the GRB resulting
from its discrete nature are substantially larger than in the starburst model.

To illustrate a potential of the GRB fluctuation analysis 
we consider the AGN model shown in Fig.~\ref{lognlogs}.
Integrated counts in this case give the GRB flux
of $\sim 1.6\times 10^{-4}$\,MeV/(s\,cm$^2$\,MeV\,sr) at 10\,MeV, what amounts
approximately to 73\,\% of the actually observed total background. Roughly
half of the model flux is produced by sources below the EGRET detection
threshold (assuming the energy index of $-2.1$). The relative rms
fluctuations produced by the Poissonian distribution of sources in the
0.26 sr bins are of the order of 5\,\% of the total (observed) GRB. However,
if we use regions comparable with the SPI field of view ($\sim 0.06$\,sr)
the rms amplitude amounts to 10\,\% and for the IBIS fov of $0.025$\,sr
it reaches 15\,\%. This figure is roughly equal to the signal-to-noise
ratio reached in the EGRET measurements in the $0.26$\,sr fields. Thus,
if the INTEGRAL instruments give similar S/N, some estimates of the GRB
fluctuations would become feasible. 

Numerical results in the above calculations depend strongly on the assumed
model parameters. In particular, if the blazars produce $100$\,\% of the GRB
(not just $73$\,\%, as assumed in the example), the amplitude
of fluctuations could be substantially larger. On the other hand,
the starburst galaxy model
provides more definite results on the amplitude of the GRB fluctuations.
The variations of the GRB predicted for this latter model are very
low and remain invariably below the
detection threshold for a wide range of model parameters.

\section{Conclusions}

Simple, discrete source models of the GRB have been constructed to investigate
effects of the background anisotropies at different angular scales.
Source counts predicted in the balazar and starburst galaxy models
are distinctly different. We have shown that precise measurements of the
GRB using of the INTEGRAL detectors would allow to distinguish between
these two cases. Isotropic GRB would indicate that the large number of faint
sources contributes to the GRB making the starburst model more likely,
while detectable GRB fluctuations would point to the blazar model.

\section*{Acknowledgments}

This paper was supported by the Polish KBN grant 2P03D~002~14.

\end{document}